\documentstyle[twoside,fleqn,npb,epsfig]{article}
\def\pipi{\pi^+ \pi^-}
\def\amu{a_\mu}
\def\amuh{a_\mu^{{\mathrm{had}}}}
\def\MZ{M_Z}

\def\az{\alpha(\MZ)}

\def\dalf{\Delta\alpha}

\def\dah{\Delta\alpha^{(5)}_{\rm had}}

\def\dahz{\Delta\alpha^{(5)}_{\rm had}(\MZ^2)}
\def\dah0{\Delta\alpha^{(5)}_{\rm had}(-s_0)}
\def\damu{\delta \amu}
\def\dalp{\delta \dalf}

\def\sfs{spectral functions}

\def\PLB{{\em Phys. Lett.}  B }

\newcommand{\be}{\begin{equation}}
\newcommand{\ee}{\end{equation}}
\newcommand{\ba}{\begin{eqnarray}}
\newcommand{\ea}{\end{eqnarray}}
\newcommand{\bea}{\begin{eqnarray*}}
\newcommand{\eea}{\end{eqnarray*}}
\newcommand{\bet}{\begin{center} \begin{tabular}}
\newcommand{\ent}{\end{tabular} \end{center}}

\newcommand{\mz}{M^2_Z}

\newcommand{\al}{\alpha}

\newcommand{\sigmmbp}{\sigma_{\mu\mu,~0 }}

\newcommand{\bary}{\begin{array}}
\newcommand{\eary}{\end{array}}
\newcommand{\noi}{\noindent}

\newcommand{\lapprox}{\raisebox{-.2ex}{$\stackrel{\textstyle<}
{\raisebox{-.6ex}[0ex][0ex]{$\sim$}}$}}
\newcommand{\crn}{\nonumber \\}
\newcommand{\nn}{\nonumber}
\newcommand{\gv}{\mbox{GeV}}

\newcommand{\bit}{\begin{itemize}}
\newcommand{\eit}{\end{itemize}}
\newcommand{\daf}{DA$\Phi$NE \ }

\newcommand{\dalh}{\Delta \alpha^{\rm had}}
\newcommand{\epm}{e^+e^-}
\newcommand{\ra}{\rightarrow}
\newcommand{\sigh}{$\sigma(\epm \ra hadrons)$}

\newcommand{\AmS}{{\protect\the\textfont2
  A\kern-.1667em\lower.5ex\hbox{M}\kern-.125emS}}

\title{%
Theoretical precision in estimates of the hadronic
contributions to $(g-2)_\mu$ and 
$\alpha_{\mbox{\footnotesize QED}}(M_Z)$\thanks{Work supported in part by 
TMR, EC-Contract No.~HPRN-CT-2002-00311 (EURIDICE)}
}

\author{F.~Jegerlehner \address{DESY, Platanenallee 6,
                        D-15738, Zeuthen, Germany},
}

\begin{document}
\onecolumn{
\renewcommand{\thefootnote}{\fnsymbol{footnote}}
\setlength{\baselineskip}{0.52cm}
\thispagestyle{empty}
\begin{flushright}
DESY 03--170 \\
October 2003\\
\end{flushright}

\setcounter{page}{0}

\mbox{}
\vspace*{\fill}
\begin{center}
{\Large\bf 
Theoretical precision in estimates of the hadronic
contributions} \\
\vspace{3mm}
{\Large\bf  to $(g-2)_\mu$ and 
$\alpha_{\mbox{\footnotesize QED}}(M_Z)$} \\

\vspace{5em}
\large
F. Jegerlehner\footnote[4]{\noindent Invited talk at Photon 2003: 
International Conference on the Structure and Interactions of the Photon 
and 15th International Workshop on Photon-Photon Collisions, 
Frascati, Italy, 7-11 Apr 2003. Work supported in part by 
TMR, EC-Contract No.~HPRN-CT-2002-00311 (EURIDICE)} 
\\
\vspace{5em}
\normalsize
{\it   DESY Zeuthen}\\
{\it   Platanenallee 6, D--15738 Zeuthen, Germany}\\
\end{center}
\vspace*{\fill}}
\newpage

\begin{abstract}
I review recent estimates of the non-perturbative hadronic vacuum
polarization contributions. Since these at present can only be
evaluated in terms of experimental data of limited precision, the
related uncertainties pose a serious limitation in our ability to make
precise predictions. Besides $\epm$-- annihilation data also $\tau$
decay spectra can help to get better predictions. Here, it is important
to account for all possible iso-spin violations in $\tau$--decay
spectra, from which $\epm$ cross sections may be obtained by an
iso-spin rotation. The observed 10\% discrepancy in the region above
the $\rho$ may be understood as a so far unaccounted iso-spin breaking
effect.
\end{abstract}

\maketitle
\section{INTRODUCTION}
Non-perturbative hadronic contributions affect electroweak precision
observables mainly via the hadronic excitations in the photon vacuum
polarization (charge screening) which leads to the energy dependence
of the effective fine structure ``constant'' $\alpha(E)$. Of particular
interest are $\alpha(M_Z)$ (precision physics at LEP/SLC)~\cite{LEP}
and $a_\mu \equiv (g-2)_\mu/2$ which has been measured at the
unbelievable precision of 0.7ppm at BNL ~\cite{BNL}.  Apart from the
electroweak effects (leptons etc.) which are calculable in
perturbation theory, a serious problem shows up for the strong
interaction effects (hadrons/quarks etc.) for the calculation of which
perturbation theory fails. Fortunately, general principles allow us to
evaluate the problematic contributions via dispersion relations from
experimental $\epm$--annihilation data represented usually in terms of
the cross section ratio 
\ba
R_\gamma(s)\equiv \frac{\sigma(e^+e^-\rightarrow \gamma^*
\rightarrow {\rm hadrons})}{ \sigma(e^+e^- \rightarrow \gamma^* \rightarrow
\mu^+ \mu^-)}\;.
\label{Rdefi}
\ea
The impact is that the errors of the experimental cross section data are
now a dominating factor for the theoretical uncertainties of electroweak
Standard Model predictions.  Therefore an art has developed of getting
precise results from measurements of often very limited precision. The
situation is also a big challenge for precision experiments on \sigh
~~as currently performed by KLOE at \daf
\cite{KLOE} and BABAR at PEP \cite{Solodov:2002xu}.  I should
mention that the measurements of $R$ are a difficult task: besides the
needed particle identification and the background rejection one has to
get
\ba
R^{\rm exp}(s)=\frac{N_{\rm had}\:{ (1+\delta_{\rm RC})}}{N_{\rm
norm}\:\varepsilon}\:{
\frac{\sigma_{\rm norm}(s)}{\sigmmbp (s)}}\; ,
\ea
where $N_{\rm had}$ is the number of observed hadronic events, $N_{\rm
norm}$ is the number of observed normalizing events, $\varepsilon$ is
the efficiency-acceptance product of hadronic events while
$\delta_{\rm RC}$ are radiative corrections to hadron production.
$\sigma_{\rm norm}(s)$ is the physical cross section for normalizing
events (including all radiative corrections integrated over the
acceptance used for the luminosity measurement) and $\sigmmbp (s)$
$=4\pi\al^2/3s$ is the normalization.  In particular this shows that a
precise measurement of $R$ requires precise knowledge of the relevant
radiative corrections.

For the normalization mostly the Bhabha process is utilized [or
$\mu\mu$ itself in some cases]. In general, it is important to be
aware of the fact that the effective fine structure constant
$\alpha(\mu)$ enters radiative correction calculations with different
scales $\mu$ in ``had'' and ``norm'' and thus must be taken into
account appropriately.

Recent advances/issues in the evaluation of the hadronic vacuum
polarization effects are based on the following results:

\noi ~~$\bullet$ Updated results from the 
precise measurements of the processes $e^+e^- \to \rho \to
\pi^+\pi^-$, $e^+e^- \to \omega \to
\pi^+\pi^-\pi^0$ and $e^+e^- \to \phi \to K_LK_S$ performed by the
CMD-2 collaboration have appeared recently~\cite{CMD2}. The
update was necessary due to an overestimate of the integrated
luminosity in the previous analysis which was published in
2002~\cite{CMD}. A more progressive error estimate (improving on
radiative corrections, in particular) allowed a reduction of the
systematic error from 1.4\% to 0.6 \% . Also some other CMD-2 and SND
data at energies $E< 1.4$ GeV have become available.

\noi ~~$\bullet$ In 2001 BES-II published their final 
$R$--data which, in the region 2.0 GeV to 5.0 GeV, allowed to reduce
the previously huge systematic errors of about 20\% to 7\%
~\cite{BES}.

\noi ~~$\bullet$ After 1997 precise $\tau$--spectral functions
became available~\cite{ALEPH,OPAL,CLEO} which, to the extent that
flavor $SU(2)_{\rm f}$ in the light hadron sector is a symmetry,
allows to obtain the iso--vector part of the $\epm$--cross
section~\cite{tsai,eidelman}. This possibility has first been
exploited in the present context in~\cite{ADH98}.

\noi ~~$\bullet$ With increasing precision of the low energy
data it more and more turned out that we are confronted with a serious
obstacle to further progress: in the region just above the
$\omega$--resonance, the iso-spin rotated $\tau$--data, after being
corrected for the known iso-spin violating effects, do not agree with
the $\epm$--data at the 10\% level~\cite{DEHZ}. Before the origin of
this discrepancy is found it will be hard to make further progress in
pinning down theoretical uncertainties.

\noi ~~$\bullet$ In this context iso-spin breaking effects in the 
relationship between the $\tau$-- and the $\epm$--data have been
extensively investigated in~\cite{CEN}. The question remains whether
all possible iso-spin violating effects have been taken into account
in which case the discrepancy would have to be attributed to
experimental problems.

\noi ~~$\bullet$ A new bound $\delta a_\mu(0.6-2.0 \gv)<0.7\:
\times 10^{-10}$~\cite{Eidelman03} for the contributions of 
$\pi\pi\gamma,\pi\eta\gamma$ which include decay products from
$\pi^0\gamma,\sigma\gamma,f\gamma,a_1\gamma$ yields a severe
constraint on possible missing contributions reported
elsewhere~\cite{Narison:2003ur}.

\noi ~~$\bullet$ New results for hadronic $\epm$ cross--sections
are expected soon from KLOE, BABAR and BELLE. These experiments,
running at fixed energies, are able to perform measurements via the
radiative return method~\cite{RR,KLOE,Solodov:2002xu}. Preliminary
results presented recently by KLOE seem to agree very well with the
final CMD-2 $\epm$--data.

\noi ~~$\bullet$ Last but not least an important change in the hadronic
contribution to $\amu$ was the change in sign of the leading hadronic
light--by--light contribution ($\pi^0$ exchange)~\cite{Knecht:2001qf}.

\noi ~~$\bullet$ Progress was made also in calculating the radiative 
corrections to $\pi^+\pi^-$ production in energy scans, for inclusive
measurements in radiative return~\cite{Hoefer:2001mx} and in photon
tagging~\cite{RR} relevant at the meson ($\Phi$,$B$) factories.

Some of these results have substantially influenced the precision of
the evaluations of the vacuum polarization effects in
$\alpha_{\mbox{\footnotesize QED}}(M_Z)$ and $(g-2)_\mu$ since
1995~\cite{EJ95}. The present status is reviewed in the following.

\section{EVALUATION OF $\alpha(M_Z)$}
The photon vacuum polarization $\Pi'_\gamma(q^2)$ modifies the fine
structure constant according to 
\ba
\;\;\;\;\alpha(q^2)\!\!&=&\frac{\alpha}{1-\Delta \alpha} \nn \\ \Delta \alpha &=&
- {\rm Re}\:\left(\Pi'_\gamma(q^2)-\Pi'_\gamma(0)\right),
\label{VPamp}
\ea

\vspace*{-4mm}

\includegraphics[scale=0.65]{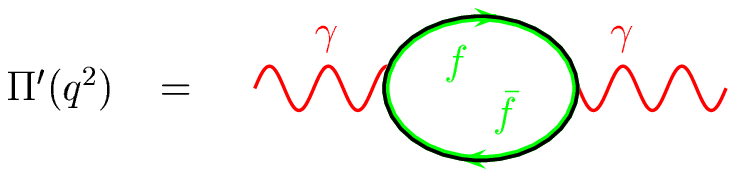}

\noi
and makes it running.
The shift $\Delta \alpha$ is large due to the large change in scale
going from zero momentum to the Z-mass scale $\mu=M_Z$ and due to the
many species of fermions contributing.

The various contributions to the shift in the fine structure constant
come from the leptons (lep = $e$, $\mu$ and $\tau$), the 5 light quarks
($u$, $b$, $s$, $c$, and $b$ and the corresponding hadrons = had) and
from the top quark:
\ba
\Delta \alpha =\Delta \alpha_{\rm lep}+\Delta^{(5)} \alpha_{\rm had}
+\Delta \alpha_{\rm top}+\cdots
\ea
Also $W$--pairs contribute at $q^2 > 2 M_W^2$ (see~\cite{JF85,FJLCN}).
The leptonic contributions are calculable in perturbation theory where
at leading order the free lepton loops yield
\footnotesize
\ba
\dalf_{{\rm lep}}(s)
& = &
      \sum\limits_{\ell=e,\mu,\tau}
      \frac{\alpha}{3\pi}
      \left[ \ln\left( s/m_\ell^2
                \right)
           - \frac{5}{3}
           + O\left( m_\ell^2/s
              \right)
      \right]    \crn
& \simeq & 0.03142 {\rm \ for \ } s=M_Z^2 \; ,
\ea
\normalsize
where $\beta_\ell = \sqrt{1 - 4m_\ell^2/s}$. This leading contribution
is affected by small electromagnetic corrections only in the next to
leading order. The leptonic contribution is actually known to three
loops~\cite{KalSab55,Ste98} at which it takes the value ($M_Z \sim
91.19$ GeV)\footnote{For $m_t \sim 174.3$ GeV we have $\Delta \alpha_{\rm
top}(M_Z^2)\simeq-\frac{\alpha}{3\pi}\frac{4}{15}\frac{M_Z^2}{m_t^2}\simeq -6 \times 10^{-5}$.}
\ba
\Delta \alpha_{\rm lep} (M_Z^2) \; \simeq \; 314.98 \: \times \: 10^{-4}.
\ea
In contrast, the corresponding free quark loop contribution gets
substantially modified by low energy strong interaction effects, which
cannot be obtained by perturbative QCD (pQCD).  As already mentioned,
fortunately, one can evaluate this hadronic term $\Delta \al _{\rm
had}^{(5)}$ from hadronic $\epm $-annihilation data by using a
dispersion relation. The relevant once subtracted vacuum polarization
amplitude (\ref{VPamp}) satisfies a convergent dispersion relation and
correspondingly the shift of the fine structure constant $\alpha$ is
given by
\ba
\Delta^{(5)} \alpha_{\rm had} &=& - \frac{\alpha s}{3\pi}\;
\bigg({\rm P}\!\!\!\!\!\!  \int_{4m_\pi^2}^{E^2_{\rm cut}} ds'
\frac{R^{\mathrm{data}}_\gamma(s')}{s'(s'-s)} \nn \\
&&~~~ + {\rm P}\!\!\!\!\!\!
\int_{E^2_{\rm cut}}^\infty ds'
\frac{R^{\mathrm{QCD}}_\gamma(s')}{s'(s'-s)}\,\, \bigg)
\label{alpint}
\ea
where
\ba
R_\gamma(s) &=& 12\pi{\rm Im}\Pi'_{\rm had}(s)
\label{Rdef}
\ea
is given by (\ref{Rdefi}).
Accordingly, the one particle irreducible (1pi) blob \\[-4mm]

\includegraphics[scale=0.65]{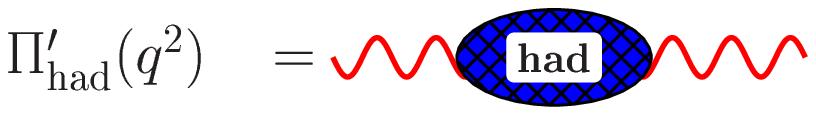}

\noi which is our relevant building block, is
given by diagrams which cannot be cut into two disconnected parts by
cutting a single photon line. At low energies it exhibits intermediate
states like
$\pi^0\gamma,\rho,\omega,\phi,\cdots,\pi\pi,3\pi,4\pi,\cdots,\pi\pi\gamma,\pi\pi
Z,$ $\cdots,\pi\pi H,\cdots,KK,\cdots$ (at least one hadron plus any
strong, electromagnetic or weak interaction contribution). The
corresponding contributions are to be calculated via a dispersion
relation from the imaginary parts which are given by the production of
the corresponding intermediate states in $\epm$--annihilation via
virtual photons (at energies sufficiently below the point where
$\gamma-Z$ interference comes into play).

A direct evaluation of the $R_\gamma (s)$--data up to
$\sqrt{s}=E_{cut}=5$ GeV and for the $\Upsilon$ resonance--region
between 9.6 and 13 GeV and applying perturbative QCD from 5.0 to 9.6 GeV
and for the high energy tail above 13 GeV at $M_Z=$ 91.19 GeV
yields\footnote{pQCD for calculating $R(s)$, as worked out to high
accuracy in Refs.~\cite{GKL}--\cite{ChHK00}, is used here only where it
has been checked to work and converge well: in non--resonant regions at
sufficiently high energies and sufficiently far from resonances and
thresholds. I have further checked that results obtained with my own
routines agree very well with the ones obtained via the recently
published program {\tt rhad-1.00}~\cite{HS02}.}:
\ba
\label{FJ03M}
\Delta \al _{\rm had}^{(5)}(\mz) &=& 0.027690 \pm 0.000353 \\
\alpha^{-1}(\mz)&=&128.922 \pm 0.049 \nn \;.
\ea
\noi The contributions from different energy ranges are shown in
Tab.~\ref{tab:dal}\footnote{Table 1 also specifies largely details of
the error handling. The different energy ranges mark typical
generations of experiments within which systematic errors are
considered to be 100\% correlated, while all errors are treated as
independent for all entries of the table.}.  The Euclidean method,
described in~\cite{EJKV98,FJ98}, allows to replace data for the Adler
function by pQCD at space--like momenta $> ~2.5~~\gv$. This yields
\ba
\Delta \al _{\rm had}^{(5)}(\mz) &=& 0.027651 \pm 0.000173 \\
\alpha^{-1}(\mz) &=&128.939 \pm 0.024 \nn \;.
\ea
with a substantially reduced error. This estimate is on a sound
theoretical basis and should not be confused with so called ``theory
driven'' estimates, which utilize pQCD in a much less controlled
manner. In future this approach would allow to evaluate $\Delta \al
_{\rm had}^{(5)}(\mz)$ with an accuracy $\dalp=0.00007$ or $0.00005$
provided future cross-section measurements allow to reduce the errors
below $\delta \sigma \:\lapprox\: 1\%$ up to $J/\psi$ or $\Upsilon$,
respectively. This assumes that in the meantime pQCD parameters will be
known with much better precision as well. This reduction of the error by
a factor about 5 is needed in order to satisfy future requirements for
precision physics at a linear collider~\cite{TDR}.
\begin{table*}[t]
\begin{tabular}{cc||r||r}
\hline
 final state &  energy range (GeV) & $\dahz$ (stat) (syst)
& $\dah0$ (stat) (syst) \\
\hline
$\chi PT$  &    (0.28, 0.32) &    0.04 ( 0.00) ( 0.00)&    0.03 ( 0.00) ( 0.00) \\
$\rho$     &    (0.28, 0.81) &   26.16 ( 0.24) ( 0.27)&   24.26 ( 0.23) ( 0.25) \\
$\omega$   &    (0.42, 0.81) &    3.02 ( 0.04) ( 0.08)&    2.75 ( 0.03) ( 0.07) \\
$\phi$     &    (1.00, 1.04) &    4.74 ( 0.07) ( 0.11)&    4.07 ( 0.06) ( 0.09) \\
$J/\psi$   &                 &   11.50 ( 0.56) ( 0.61)&    4.06 ( 0.19) ( 0.19) \\
$\Upsilon$ &                 &    1.27 ( 0.05) ( 0.07)&    0.07 ( 0.00) ( 0.00) \\
    had&    (0.81, 1.40) &   12.92 ( 0.13) ( 0.52)&   11.05 ( 0.11) ( 0.43) \\
    had&    (1.40, 3.10) &   27.13 ( 0.11) ( 0.60)&   15.75 ( 0.06) ( 0.37) \\
    had&    (3.10, 3.60) &    5.31 ( 0.11) ( 0.10)&    1.90 ( 0.04) ( 0.04) \\
    had&    (3.60, 9.46) &   51.49 ( 0.25) ( 3.00)&    8.41 ( 0.04) ( 0.44) \\
    had&    (9.46,13.00) &   18.59 ( 0.25) ( 1.36)&    0.90 ( 0.01) ( 0.07) \\
    pert& (13.0,$\infty$) &  115.59 ( 0.00) ( 0.12)&    1.09 ( 0.00) ( 0.00) \\
\hline
    data   &    (0.28,13.00) &  162.14 ( 0.74) ( 3.46)&     73.21 ( 0.33) ( 0.80) \\
    total  &                 &  277.73 ( 0.74) ( 3.46)&     74.30 ( 0.33) ( 0.80) \\
\hline
\end{tabular}
\caption{%
Results for $\dahz^{\mathrm{data}}$ and
$\dah0^{\mathrm{data}}$ ($\sqrt{s_0}=2.5$ GeV).}
\label{tab:dal}
\end{table*}
Our analysis is as close to the experimental results as possible by
utilizing the trapezoidal rule together with PDG rules for taking
weighted averages between different experiments as described in detail
in~\cite{EJ95}.

\begin{figure}[h]
\begin{picture}(120,60)(15,27)
\includegraphics[scale=0.45]{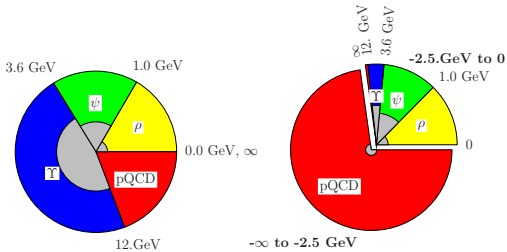}
\end{picture}

\caption{Comparison of the distribution of contributions and errors
(shaded areas scaled up by 10) in the standard (left) and the Adler
function based approach (right), respectively.}
\label{fig:alpsta}
\end{figure}

The most important ingredient of our analysis are the $\epm$--data
which we described in detail in~\cite{EJ95} (see also~\cite{ADH98})
and the new data which have become available since
then~\cite{Jegerlehner:2003ip}. The distribution of hadronic
contributions to $\dalh$ in the $\epm$--data based approach in shown
in Fig.~\ref{fig:alpsta}.

\vspace*{-3mm}

Our results are in good agreement with other recent analyzes
\small
\ba
\Delta \al _{\rm had}^{(5)}(\mz)  
 = \left\{ \begin{tabular}{ll}
   {$0.027680 \pm 0.000360$} &{\cite{BP01}}\\
   {$0.027690 \pm 0.000180$} &{\cite{HMNT}}
\end{tabular} \right.
\ea
\normalsize
The corresponding values for the effective fine structure constant
are:
\small
\ba
\alpha^{-1}(\mz) = \left\{
\begin{tabular}{ll}
  {$128.935 \pm 0.049$ } &{\cite{BP01}}\\
  {$128.933 \pm 0.025$ } &{\cite{HMNT}}
\end{tabular} \right.
\ea
\normalsize

The numerical agreement does not necessarily mean agreement between
the different approaches.

\section{EVALUATION OF $a_\mu \equiv (g-2)_\mu/2$}

The anomalous magnetic moment of the muon $\amu$ provides one of the
most precise tests of the quantum field theory structure of QED and
indirectly at a deeper level also of the electroweak SM. The precision
measurement of $a_\mu$ is a very specific test of the magnetic helicity
flip transition $\bar{\psi}_L\:{\sigma_{\mu \nu}}\: {
F^{\mu\nu}}\:\psi_R$, a dimension 5 operator which is forbidden for any
species of fermions at the tree level of any renormalizable theory. In
the SM it is thus a finite prediction which can be tested unambiguously
to the extent that we are able to calculate it with the necessary
accuracy. For the perturbative part of the SM an impressive precision
has been reached.  Excitingly the new experimental result from
Brookhaven~\cite{BNL} which reached a substantial improvement in
precision shows a 1.9[0.7] $\sigma$ deviation from the theoretical
prediction: $\left| a_\mu^{\rm exp}-a_\mu^{\rm
the}\right|=221(113)[074(104)] \times 10^{-11}$, depending on whether
one trusts more in an $e^+e^-$--data[$\tau$--data] based evaluation of
the hadronic vacuum polarization contribution~\cite{DEHZ}.

Again contributions from virtual creation and reabsorption of strongly
interacting particles cannot be computed with the help of pQCD and
cause serious problems. Fortunately the major such contribution again
enters via the photon vacuum polarization which can be calculated
along the lines discussed for the effective charge. The contribution
is described by the diagram\\[-1mm]

\begin{figure}[h]
\begin{picture}(120,60)(15,35)
\includegraphics[scale=0.65]{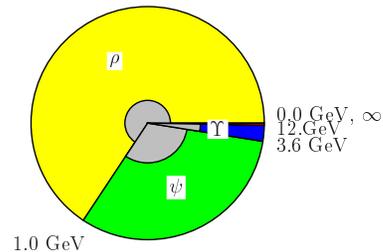}
\end{picture}
\caption{The distribution of contributions and errors
(shaded areas scaled up by 10) for $\amuh\;$. }
\label{fig:gmusta}
\end{figure}

\centerline{%
\includegraphics[scale=0.65]{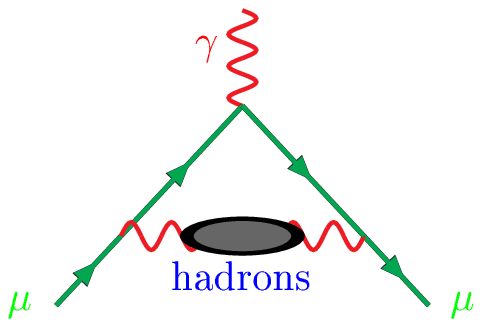}}

\noindent
and is represented by the integral
\ba
\amuh &=& \left(\frac{\alpha m_\mu}{3\pi}
\right)^2 \bigg(
\int\limits_{4 m_\pi^2}^{E^2_{\rm cut}}ds\,
\frac{R^{\mathrm{data}}_\gamma(s)\;\hat{K}(s)}{s^2}  \nn \\
&&~~~ + \int\limits_{E^2_{\rm cut}}^{\infty}ds\,
\frac{R^{\mathrm{pQCD}}_\gamma(s)\;\hat{K}(s)}{s^2}\,\, \bigg)
\label{AM}
\ea
which is similar to the integral (\ref{alpint}), however with a different kernel
$K(s)$ which may conveniently be written in terms of the variable
\bea
x=\frac{1-\beta_\mu}{1+\beta_\mu}\;,\;\;\beta_\mu=\sqrt{1-4m^2_\mu/s}
\eea
and is given by \footnotesize
\ba
K(s)&=&\frac{x^2}{2}\:(2-x^2) 
    +\frac{(1+x^2)(1+x)^2}{x^2}\crn && \!\!\!
       \left(\ln(1+x)-x+\frac{x^2}{2} \right) 
    +\frac{(1+x)}{(1-x)}\:x^2 \ln(x).
\label{KS}
\ea \normalsize
The integral (\ref{AM}) is written in terms of the rescaled
function \bea \hat{K}(s)=\frac{3 s}{m^2_\mu}K(s) \eea which is
bounded: it increases monotonically from 0.63 at threshold
$s=4m^2_\pi$ to 1 at $\infty$. Note the extra $1/s$--enhancement of
contributions from low energies in $\amu$ as compared to $\Delta
\alpha$.
\small
\begin{table}[t]
\begin{tabular}{cc||r}
\hline
 final state &  range (GeV) & $\damu$ (stat) (syst)\\
\hline
$\chi PT$  &    (0.28, 0.32) &   2.14 ( 0.02) ( 0.03)\\
$\rho$     &    (0.28, 0.81) & 429.02 ( 4.95) ( 5.59) \\
$\omega$   &    (0.42, 0.81) &  37.99 ( 0.46) ( 1.03) \\
$\phi$     &    (1.00, 1.04) &  36.07 ( 0.50) ( 0.83) \\
$J/\psi$   &                 &   8.74 ( 0.41) ( 0.40) \\
$\Upsilon$ &                 &   0.11 ( 0.00) ( 0.01) \\
    had&    (0.81, 1.40) & 105.17 ( 1.18) ( 3.29) \\
    had&    (1.40, 3.10) &  56.33 ( 0.21) ( 1.47) \\
    had&    (3.10, 3.60) &   4.06 ( 0.08) ( 0.08) \\
    had&    (3.60, 9.46) &  14.43 ( 0.07) ( 0.75) \\
    had&    (9.46,13.00) &   1.30 ( 0.02) ( 0.10) \\
    pQCD& (13.0,$\infty$) &   1.53 ( 0.00) ( 0.00) \\
\hline
    data   &    (0.28,13.00) & 693.22 ( 5.15) (6.83) \\
    total  &                 & 694.75 ( 5.15) (6.83) \\
\hline
\end{tabular}
\caption{Results for $\damu^{\mathrm{data}}$.}
\label{tab:dam}
\end{table}
\normalsize
The relative importance of various regions is illustrated in
Tab.~\ref{tab:dam} and Fig.~\ref{fig:gmusta}. The update of the
results~\cite{EJ95}, including the recent data from CMD-2 and BES-II
yields \ba \amuh &=& (694.75 \pm 8.56) \times 10^{-10}\;\;.
\label{eq:amuhad} 
\ea 
The most recent BNL $(g_\mu-2)$
measurement~\cite{BNL} gives (world average)
\bea a_\mu^{\rm exp} &=& {(11659203~~ \pm
~8~~)} \times 10^{-10}~ \mbox{ } \eea which compares with
the theoretical prediction\footnote{Recent new results concern the
hadronic light-by-light contribution~\cite{Andreas} and the
$O(\alpha^4)$ QED contribution to $a_e$~\cite{Kino02}.} \bea a_\mu^{\rm
the} &=& {(11659169.6 \pm ~9.4)} \times 10^{-10}~~ \mbox{
(SM)}\;. ~~~~~~~ 
\eea

\newpage

\begin{figure}[h]
\begin{picture}(120,60)(-15,35)
\includegraphics[scale=0.5]{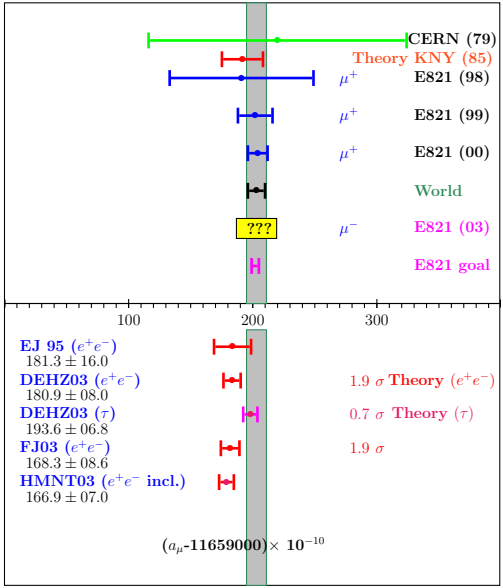}
\end{picture}

\vspace*{1.0cm}

\caption{Experimental (upper part) and theoretical (lower part) status 
of $\amu$. }
\label{fig:gm2status}
\end{figure}

The new analysis~\cite{DEHZ} is ``data--driven'' like~\cite{EJ95,ADH98}
and confirms a substantial discrepancy between $\epm$-- and
$\tau$--data. The $\tau$--based result agrees with the corresponding
result of~\cite{ADH98}.  The status is illustrated in
Fig.~\ref{fig:gm2status}.  We refer to Ref.~\cite{Nyffeler:2003vb} for a
recent review and possible implications.

\vspace*{3.8cm}

\vspace*{-6mm}

\section{$e^+e^-$ CROSS-SECTIONS VIA \hfill \\ $\tau$--DECAY 
SPECTRAL FUNCTIONS \hfill }
A substantial improvement of the evaluation of $\amuh$ would be possible,
by including the $\tau$--data, provided one would understand iso--spin
viola\-ting effects sufficiently well~\cite{CEN}. This has been pioneered by
Ref.~\cite{ADH98}. Here one utilizes the fact that the vector--current
hadronic $\tau$--decay spectral functions are related to the
iso--vector part of the $\epm$--annihilation cross--section via an
iso-spin rotation: $$ \tau^- \ra X^-
\nu_\tau\;\;\; \leftrightarrow \;\;\; e^+ e¯ \ra X^0$$ where $X^-$ and
$X^0$ are related hadronic states. 
The precise relationship may be derived by comparing diagrams like:

\includegraphics[scale=0.65]{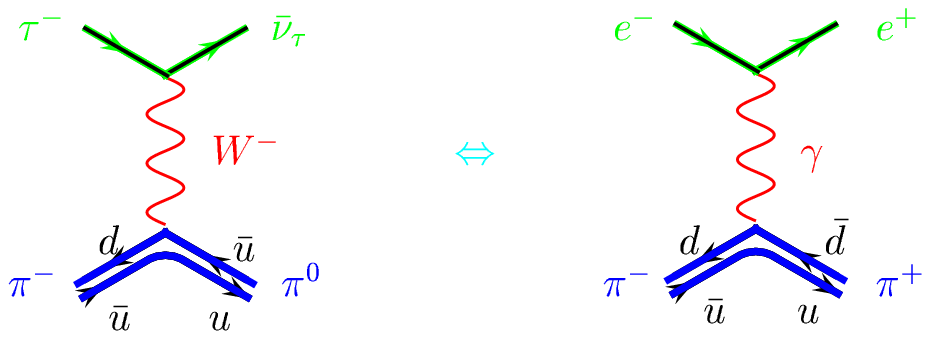}

\noi
which for the $\epm$ case translates into
\begin{equation}
\sigma^{(0)}_{\pi\pi}\equiv 
\sigma_0 (e^+e^- \to \pi^+\pi^-) = \frac{4\pi\alpha^2}{s}\:\times  v_0(s)
\label{eepp}
\end{equation}
and for the $\tau$ case into
\ba
\frac{1}{\Gamma} \frac{d\Gamma}{ds}
 (\tau^- \to \pi^-\pi^0 \nu_\tau) = 
\frac {6\pi |V_{ud}|^2 S_{EW}}{m_\tau^2}\times v_-(s) 
\times \!\!\!\!\!\!\!\!\!\!\!\! \nn \\ 
\frac{B(\tau^-\rightarrow \nu_\tau\,e^-\,\bar{\nu}_e)}
{B(\tau^-\rightarrow \nu_\tau\,\pi^- \pi^0)}\: 
\left(1-\frac{s}{m_\tau^2}\right)
\left(1 + \frac {2s}{m_\tau^2}\right)\: 
\label{taupp}
\ea
where $|V_{ud}|=0.9752\pm0.0007$~\cite{PDG02} denotes the CKM weak
mixing matrix element and $S_{\mathrm{EW(new)}}=1.0233\pm0.0006$
[$S_{\mathrm{EW(old)}}=1.0194$] accounts for electroweak radiative
corrections~\cite{Marciano:vm,Braaten:1990ef,Decker:1994ea,CEN}.  The \sfs\
are obtained from the corresponding invariant mass distributions. The
$B(i)$'s are branching ratios. SU(2) symmetry (CVC) would imply
\begin{equation}
v_-(s) =  v_0(s)\;\;.
\label{CVCrel}
\end{equation}
The spectral functions $v_i(s)$ are related to the pion form factors
$F^i_\pi(s)$ by
\begin{equation}
v_i(s)=\frac{\beta_i^3(s)}{12\pi} |F^i_\pi(s)|^2\;\;;\;\;\;(i=0,-)
\label{sfvsff}
\end{equation}
where $\beta_i^3(s)$ is the pion velocity. The difference in phase
space of the pion pairs gives rise to the relative factor
$\beta^3_{\pi^-\pi^0}/\beta^3_{\pi^-\pi^+}$.

With the precision of the validity of CVC, thus the $\tau$--data allow
us to improve the $I=1$ part of the $\epm$ cross--section which by
itself is not a directly measurable quantity. It mainly improves the
knowledge of the $\pipi$ channel ($\rho$--resonance contribution)
which is dominating in $\amuh$ (72\%). After taking into account the
known iso-spin breaking effects~\cite{CEN} the $\tau$--data show
substantial discrepancies in comparison with the $\epm$--data (about
10\% just above the $\rho$--resonance). This issue can certainly be
settled by the radiative return experiments with KLOE~\cite{KLOE} at
LNF/Frascati and with BABAR~\cite{Solodov:2002xu} at SLAC. In fact
preliminary results from KLOE are close to the CMD-2 results. At
present one obtains incompatible predictions for $\amuh$ based on
$\epm$--data or on $\tau$--data (see Fig.~\ref{fig:gm2status}).
\section{ISO-SPIN BREAKING \hfill \\ CORRECTIONS IN $\tau$
VS. $e^+e^-$ \hfill}

\vspace*{0.5cm}

\begin{figure}[h]
\begin{picture}(120,60)(-15,35)
\includegraphics[scale=0.5]{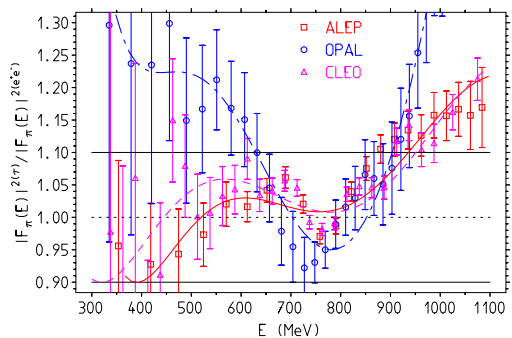}
\end{picture}

\vspace*{0.5cm}

\caption{The ratio between $\tau$--data sets from ALEPH, OPAL and CLEO 
and the $I=1$ part of the CMD-2 fit of the $\epm$--data. The curves
which should guide the eye are fits of the ratios using 8th order
Tschebycheff polynomials.}
\label{fig:ratio}
\end{figure}
Before a precise comparison is possible all kind of iso-spin breaking
effects have to be taken into account. As mentioned earlier, this has
been investigated in~\cite{CEN} for the most relevant $\pi \pi$
channel. For the $\tau$ version of the pion form
factor, following from (\ref{taupp}) and (\ref{sfvsff}), we perform
the iso--spin breaking corrections
\be
r_{\mathrm{IB}}(s)=\frac{1}{G_{\rm EM}(s)} \:
\frac{\beta^3_{\pi^-\pi^+}}{\beta^3_{\pi^- \pi^0}} 
\:\frac{S_{\mathrm{EW(old)}}}{S_{\mathrm{EW(new)}}}
\label{rib}
\ee
with $G_{\rm EM}(s)$ (from~\cite{CEN}) accounting for the QED
corrections of the $\tau$--decays. Final state radiation (FSR) is
modeled by scalar QED. The resulting bare form factor
$|F_\pi|^{2\:{I=1}}_{(\tau)}(s)$ compares to the bare (vacuum
polarization and FSR subtracted) form factor
$|F_\pi|^{2\:{I=1}}_{(\epm)}(s)$, which can be obtained from the
measured pion form factor $|F_\pi|^{2\:{\rm exp}}_{(\epm)}(s)$ by 
subtracting the $\rho - \omega$ mixing effects:  \small
\ba
|F_\pi|^{2\:{I=1}}(s) = |F_\pi|^{2\:{\rm exp}}(s) /
|\left(1+ \frac{\epsilon  s}{(s_\omega-s)}\right)|^2
\ea \normalsize
with $s_\omega = (M_\omega - \frac{i}{2} \Gamma_\omega)^2$,
$\epsilon$ determined by a fit to the data: $\epsilon = 0.00172$.
In Fig.~\ref{fig:ratio} we display $|F_\pi|^{2\:{I=1}}_{(\tau)}(s)
/|F_\pi|^{2\:{I=1}}_{(\epm)}(s)$ which shows large deviations from the
CVC line represented by unity.

These above corrections were applied also in~\cite{DEHZ} and revealed
that they were not sufficient to remove the unexpectedly large
discrepancy (see ~\cite{DEHZ} for details). The only large
effect I am aware of (of order 10\%) which is in the game of the
comparison is a possible shift of the invariant mass of the pion-pairs
in the $\rho$ resonance region. An idea one gets if one is looking at
the experimental $\rho$--mass values, shown in the particle data
tables~\cite{PDG02} (``dipole shape'').  If the energy calibration of
the $\pi\pi$--system would be too low in $\epm$ measurements or to high
in $\tau$ measurements by 1\% one could easily get a 10\% decrease or
increase in the tail, respectively. Since the $\rho^\pm-\rho^0$ mass
difference as well as the difference in the widths
$\Gamma^{\pm,0}(\rho \to
\pi\pi,\pi\pi\gamma)$ are neither experimentally nor theoretically
established, corresponding iso-spin violations cannot be corrected for
appropriately. Note that the subtraction of the large and strongly
energy dependent vacuum polarization effects necessary for the
$\epm$--data, which seems to worsen the problem, is properly treated
in the analysis.

\section{EVALUATION OF $a_\mu$ VIA THE \hfill \\ ADLER FUNCTION \hfill}

In Ref.~\cite{EJKV98} it has been shown how one can obtain a better
control on the validity of pQCD by utilizing analyticity and looking
at the problem in the $t$--channel (Euclidean field theory
approach). It has been found that ``data'' may be safely replaced by
pQCD at $\sqrt{-t} \geq 2.5 \gv$. An application to the calculation of
the running fine structure constant has been discussed
in~\cite{FJ98,Jegerlehner:2003ip}. Here we consider the application to the
calculation of $\amuh$. Starting point is the basic integral
representation \small
\be
\amuh=\frac{\alpha}{\pi}\int\limits_0^\infty\frac{ds}{s}
\int\limits_0^1 dx\:\frac{x^2\:(1-x)}{x^2+(1-x)\:s/m^2_\mu}\:
\frac{\alpha}{3\pi} \: R(s).  
\ee \normalsize
If we first integrate over $x$ we find the well known standard
representation as an integral along the cut of the vacuum polarization
amplitude in the time--like region, while an interchange of the order
of integrations yields an integral over the hadronic shift of the fine
structure constant in the space--like domain~\cite{LPdR72}:
\be
\amuh=\frac{\alpha}{\pi}\int\limits_0^1 dx\:(1-x)\: \dalh
\left(-Q^2(x)\right)
\label{RAI}
\ee
where $Q^2(x)\equiv \frac{x^2}{1-x}m_\mu^2$ is the space--like square
momentum--transfer or
$$x=\frac{Q^2}{2m_\mu^2}\:\left(\sqrt{1+\frac{4m_\mu^2}{Q^2}}-1\right)\;\;.$$
In this approach we (i) calculate the Adler function from the
$\epm$--data and pQCD for the tail above 13 GeV, (ii) calculate the
shift $\dalh$ in the Euclidean region with or without an additional
cut in the $t$--channel at 2.5 GeV and (iii) calculate $\amuh$
via~(\ref{RAI}).\\

Alternatively, by 
performing a partial integration in (\ref{RAI}) one finds
\be
\amuh=\frac{\alpha}{\pi}m_\mu^2 \int\limits_0^1 dx\:x\:(2-x)\:
\left(D(Q^2(x))/Q^2(x)\right)
\label{ADI}
\ee 
by means of which the number of integrations may be reduced by
one. The evaluation in both forms provides a good stability test of
the numerical integrations involved.

Utilizing the most recent $\epm$--data we obtain a result which agrees
with the values obtained by the direct evaluation also in the
error. Not too surprisingly, as is well known, the contribution to
$\amuh$ is dominated by the low energy $\epm$--data below 1 GeV; here
the replacement of data by pQCD does not reduce the uncertainty. The
reason is hat the pQCD contribution replacing the Euclidean Adler
function at $\sqrt{-t}> 2.5 \gv $ shows a substantial uncertainty due
to the uncertainty of the charm mass $m_c(m_c)=1.15...1.35\;\gv$. The
uncertainty in the strong coupling constant
$\alpha_s(M_Z^2)=0.120\pm0.003$ is small and is not the dominating
effect. In contrast to other authors which use pQCD for estimating
$R(s)$ directly, we do not obtain a reduction of the error. Of course
our cut at 2.5 GeV, which we think is all we can justify, is more
conservative than the 1.8 GeV in the time--like region anticipated
elsewhere. Thus the best value we can obtain from presently available
$\epm$--data alone is the result (\ref{eq:amuhad}). The Euclidean method
of calculating $\amuh$ will only be useful once the QCD parameters
will be determined much more accurately. For recent progress in this
direction we refer 
to~\cite{Kuhn:2001dm,DellaMorte:2002vm,Rolf:2002gu,Lellouch:2002nj}.

\section{SUMMARY AND OUTLOOK}
Evaluations of the hadronic vacuum polarization effects based on
$\epm$--data agree reasonably well between different groups, especially
for the ``conservative'' analyzes which rely directly on the
experimental data. The accuracy typically is 1.3\%. Future precision
physics at a high luminosity linear collider would require an
improvement of the precision in $\az$ by a factor of about 5 at
least. The ``theory-driven'' analyzes which replace data by
perturbatively calculated $R$--ratios obtain much smaller errors (about
a factor 2) in general. Obviously much more reliable is the Adler
function ``monitored'' evaluation, which utilizes pQCD only for the
smooth Adler function which is much better accessible to perturbation
theory at sufficiently large space-like momenta ($|Q|> 2.5 \gv$). The
reason why this approach is not so popular is the fact that the method
is much more elaborate. Necessary improvements are possible only by
measurements at the 1\% level of the hadronic cross-sections up to
$J/\psi$ or better up to the $\Upsilon$. Plans for future experiments in
this direction exist at many places (Novosibirsk, Frascati, SLAC,
Beijing, Cornell and KEK). The $\tau$--data are potentially important
for improving the $\amuh$ evaluation. However, they can only be utilized
after appropriate iso-spin corrections. It is likely that the observed
$\tau$ vs $\epm$ disagreement is a so far unaccounted iso-spin breaking
effect, due to the difference in the physical mass and width of the
charged $\rho^\pm$ observed in $\tau$--decays and of the neutral
$\rho^0$ seen in $\epm$ annihilation.\\[-3mm]

{\bf Acknowledgments\\[-3mm]~}

It thank the organizers of the Conference Photon 2003 and in particular
Giulia Pancheri for the kind invitation and the excellent hospitality
at Frascati. Thanks also to Jochem Fleischer for carefully reading the 
manuscript.\\[-3mm]


\end{document}